# The Monitoring System of the End-Cap Calorimeter in the Belle II experiment

V. Izzo, A. Aloisio, F. Ameli, A. Anastasio, P. Branchini, F. Di Capua, R. Giordano, A. Kuzmin, K. Miyabayashi, I. Nakamura, M. Nakao, G. Tortone and S. Uehara

*Abstract*–The Belle II experiment is presently in phase-2 operation at the SuperKEKB electron-positron collider in KEK (Tsukuba, Japan). The detector is an upgrade of the Belle experiment at the KEKB collider and it is optimized for the study of rare B decays, being also sensitive to signals of New Physics beyond the Standard Model.

The Electromagnetic Calorimeter (ECL) is based on CsI(Tl) scintillation crystals. It splits in a barrel and two annular end-cap regions, these latter named Forward and Backward, according to the asymmetric design of the collider.

CsI(Tl) crystals deliver a high light output at an affordable cost, however their yield changes with temperature and can be permanently damaged by humidity, due to the strong chemical affinity for moisture. Each ECL region is then equipped with thermistors and humidity probes to monitor environmental data. While sensors and cabling have been inherited from the original Belle design, the ECL monitoring system has been fully redesigned.

In this paper, we present hardware and software architecture deployed for the 2112 CsI(Tl) crystals arranged in the Forward and Backward end-caps. Single-Board Computers (SBCs) have been designed ad-hoc for embedded applications. For sensor read-out, a data-acquisition system based on 24-bit ADCs with local processing capability has been realized and interfaced with the SBCs. EPICS applications send data across the Local Area Network for remote control and display.

## I. INTRODUCTION

THE Belle II experiment [1] is presently in phase-2 operation at the SuperKEKB electron-positron collider in KEK (Tsukuba, Japan). The Belle II detector is a major upgrade of the Belle experiment, which operated in the previous years at the KEKB collider. The Bell II detector is optimized for the study of rare B decays, but it also sensitive to signals of New Physics beyond the Standard Model, including studies of the dark sector.

The Belle II Electromagnetic Calorimeter (ECL) [2] is based on CsI(Tl) scintillation crystals. It is partitioned in a central (barrel) region and two end-cap regions. The two end-cap regions are named Forward and Backward, following the asymmetric design of the collider.

CsI(Tl) crystals provide a very high light yield, at a reasonable cost, but they are very sensitive to temperature and humidity: their light yield changes with temperature and they are highly hygroscopic, so that they can be permanently damaged by humidity [3][4].

Each ECL region is then equipped with thermistors and humidity probes, in order to monitor environmental data. The ECL monitoring system has been fully redesigned in the recent years, with the constraints that sensors and cabling have been inherited from the original Belle design.

In this paper, we present hardware and software architecture deployed for the 2112 CsI(Tl) crystals arranged in the Forward and Backward end-caps. Single-Board Computers (SBCs) have been designed *ad-hoc* for such embedded application [5]. For sensor read-out, a high-performance data-acquisition system based on 24-bit ADCs with local processing capability has been designed and implemented; it is interfaced with the SBCs. EPICS applications send data across the Local Area Network for remote control and display.

## II. THE MONITORING SYSTEM

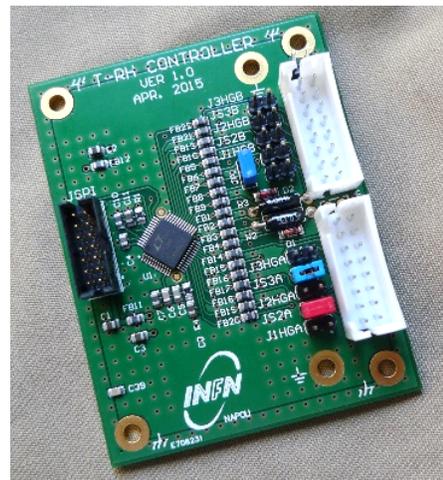

Fig. 1. The Temperature and Relative Humidity (T-RH) controller for the ECL end-cap monitoring system.

The core of this monitoring system is based on the LTC2983 System-on-Chip [6]. This device allows implementing a complete Digital Temperature Measurement System, including three 24-bit ADCs, analog multiplexing, current sources for sensor excitation, microprocessor assisted linearization. A Controller board (Fig.1) has been designed to handle two sectors of an end-cap.

A pair of Controllers are read out via two galvanically isolated SPI serial busses by a SBC based on ARM Cortex-10 microprocessor, running LINUX. SBC and Controllers are housed on a Eurocard 6U carrier board which is enclosed in a shielded box. Four boards are needed to read-out one end-cap.

The low-noise and shielded design preserves at system level the 24-bit accuracy of the LTC2983. The noise floor of a channel with grounded input has been measured in-the-field to be 0.7µV (Fig. 2).

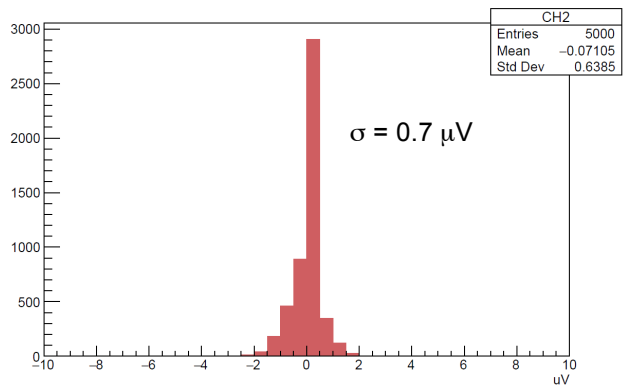

Fig. 2. The T-RH Controller Noise floor (grounded input).

Common and Normal Mode Rejection Ratios have been measured too, as well as the effectiveness of the 50Hz and 60Hz noise filters available in the ΔΣ ADCs. All the system tests have shown results in excellent agreement with the component datasheet. The temperature trend recorded by the system is published on a web page (Fig. 3). The integration in the slow control EPICS framework of the Belle II experiment is presently ongoing.

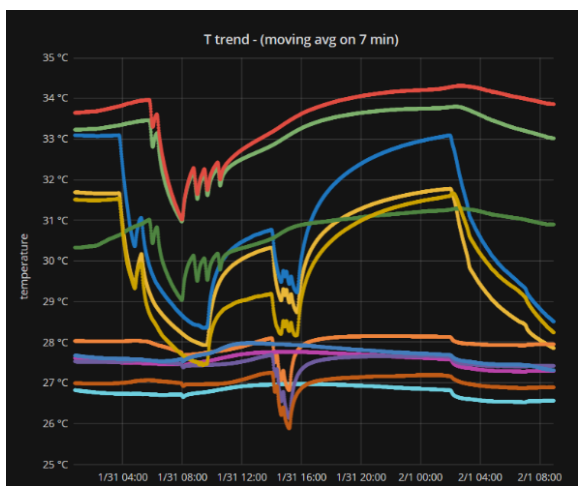

Fig. 3. Temperature trend in the forward endcap.

### III. CONCLUSIONS

The monitoring system of the Electromagnetic Calorimeter of the Belle II experiment has been successfully tested in the lab.

The system is based on the LTC2983 System-on-Chip and we achieved an excellent performance, which rivals the one of lab-grade equipment. The slow control software is presently being integrated in the framework of the experiment.